\documentclass{emulateapj}

\slugcomment{Accepted by ApJL on 2010 March 31}

\shorttitle{Parsec-Scale Localization of SDSS J1536+0441A}
\shortauthors{Wrobel \& Laor}

\begin{document}

\title{Parsec-Scale Localization of the Quasar SDSS J1536+0441A,
       a Candidate Binary Black Hole System}

\author{J. M. Wrobel\altaffilmark{1} and A. Laor\altaffilmark{2}}

\altaffiltext{1}{National Radio Astronomy Observatory, P.O. Box O,
Socorro, NM 87801; jwrobel@nrao.edu}
\altaffiltext{2}{Physics Department, Technion, Haifa 32000, Israel;
laor@physics.technion.ac.il}

\begin{abstract}
The radio-quiet quasar SDSS J1536+0441A shows two broad-line emission
systems, recently interpreted as a binary black hole (BBH) system with
a subparsec separation; as a double-peaked emitter; or as both types
of systems.  The NRAO VLBA was used to search for 8.4 GHz emission
from SDSS J1536+0441A, focusing on the optical localization region for
the broad-line emission, of area 5400 mas$^2$ (0.15 kpc$^2$).  One
source was detected, with a diameter of less than 1.63 mas (8.5 pc)
and a brightness temperature $T_b > 1.2 \times 10^7$ K.  New NRAO VLA
photometry at 22.5 GHz, and earlier photometry at 8.5 GHz, gives a
rising spectral slope of $\alpha = 0.35\pm0.08$.  The slope implies an
optically thick synchrotron source, with a radius of about 0.04 pc,
and thus $T_b \sim 5 \times 10^{10}$ K.  The implied radio-sphere at
rest frame 31.2~GHz has a radius of 800 gravitational radii, just
below the size of the broad line region in this object.  Observations
at higher frequencies can probe whether or not the radio-sphere is as
compact as expected from the coronal framework for the radio emission
of radio-quiet quasars.
\end{abstract}


\section{Motivation}\label{motivation}
Binary black hole (BBH) systems with subparsec scales are predicted in
merging scenarios for galaxy evolution and also factor prominently in
predictions for the gravitational wave background
\citep[e.g.,][]{col09}.  But can such BBH systems be found?  A
promising search method - identifying candidate BBH systems through
their optical broad-line emission properties - recently yielded three
such candidates: SDSS J153636.22+044127.0 \citep{bor09}, SDSS
J092712.65+294344.0 \citep{bog09,dot09}, and SDSS J105041.35+345631.3
\citep{shi09}.

In this Letter we focus on SDSS J153636.22+044127.0 (SDSS J1536+0441
hereafter), a quasar at a redshift $z = 0.388$ \citep{bor09}.  For the
assumed flat cosmology\footnote{$H_0 = 71$~km~s$^{-1}$~Mpc$^{-1}$ and
  $\Omega_m = 0.27$, implying a luminosity distance of 2.1 Gpc, an
  angular size distance of 1.1 Gpc and a scale of 5.2 pc per mas.},
the inferred 0.1 pc separation subtends 0.02 mas \citep{bor09}.  This
quasar has also been interpreted as a lone double-peaked emitter
\citep[DPE;][]{gas10,cho10} or as twin DPEs with a subparsec
separation \citep{tan09}.  The emission lines from a DPE are thought
to arise from rotational motion in a relativistic accretion disk.
SDSS J1536+0441 is radio quiet and our imaging of it at 8.5 GHz
\citep{wro09} revealed two faint sources, SDSS J1536+0441A and
J1536+0441B, separated by 0.97\arcsec\, (5.1 kpc) with each source
being unresolved with a diameter of less than 0.37\arcsec\, (1.9 kpc).
It is now clear that SDSS J1536+0441A is hosted by the radio-quiet
quasar (RQQ), while SDSS J1536+0441B is hosted by a companion
radio-loud elliptical galaxy \citep{dec09,lau09}.

In the BBH scenarios for SDSS J1536+0441A \citep{bor09,tan09}, each of
the two broad-line emission systems is itself a quasar.  Recent
optical spectroscopy finds that the broad-line emission systems have
relative positions that localize them to 3 $\sigma$ bands of full
width 90 mas (470 pc) along a position angle (PA) of 48\arcdeg\,
\citep{cho10} and 60 mas (310 pc) along PA 90\arcdeg\, (T. Boroson,
2010, private communication).  The resulting parallelogram localizes
SDSS J1536+0441A to an area of 5400 mas$^2$ (0.15 kpc$^2$).
Improvements in the localization of the emission from SDSS J1536+0441A
would further test its candidacy as a BBH system.  In this Letter we
use measurement techniques at radio frequencies to investigate the
compactness of SDSS J1536+0441A, first by seeking evidence for a
synchrotron self-absorbed spectrum and then by direct imaging with mas
resolution to localize the emission on parsec scales.  Our new imaging
is presented in \S~\ref{imaging} and its implications are explored in
\S~\ref{implications}.  A summary appears in \S~\ref{summary}.

\section{New Imaging}\label{imaging}

The CnB configuration of the NRAO VLA\footnote{Operated by the
  National Radio Astronomy Observatory, which is a facility of the
  National Science Foundation, operated under cooperative agreement by
  Associated Universities, Inc.} \citep{tho80} was used under proposal
code AL739 to observe SDSS J1536+0441A and J1536+0441B for 2 hours
near transit on UT 2009 May 27.  We followed the strategies described
by \citet{wro09}, except that for our new observations the center
frequency was 22.5~GHz, the switching time was 120~s, and the
amplitude scale was set to an accuracy of about 5\%.  The elliptical
Gaussian resolution, 0.96\arcsec\, times 0.81\arcsec\, at PA
-89\arcdeg, was sufficient to obtain photometry for each source.  We
will report elsewhere on SDSS J1536+0441B.  For SDSS J1536+0441A, the
flux density was $S = 1.65 \pm 0.11$~mJy and comparison with the 8.5
GHz value \citep{wro09} implies a rising spectrum with index $\alpha =
0.35\pm0.08$ ($S \propto \nu^\alpha$).  The 2009 December 31 release
of the NRAO AIPS software was used for calibration and imaging.

The NRAO VLBA$^4$ \citep{nap94} was used for 5 hours under proposal
code BL168 on 2009 October 14 UT to search for 8.4 GHz emission from
SDSS J1536+0441A.  Phase-referenced observations were made in the
nodding style, using 32 MHz per circular polarization.  A 120~s scan
of SDSS J1536+0441A was preceded and followed by a 60~s scan of the
reference source J1539+0430, favorably located at a switching angle of
0.7\arcdeg\, from SDSS J1536+0441A.  The 2009 December 31 release of
the NRAO AIPS software was used for calibration and imaging.  We
followed the calibration strategies described by \citet{wro06} and the
large-area imaging strategies described by \citet{wro05}.

We conducted a large-area VLBA search in a region set by the 3
$\sigma$ VLA astrometric accuracy \citep{wro09} and of area 280,000
mas$^2$.  Using natural weighting of the visibility data, this search
achieved an elliptical Gaussian resolution of 2.57 mas times 1.04 mas
at PA -5.2\arcdeg, a geometric-mean resolution of 1.63 mas, and a
root-mean-square sensitivity of 0.056 mJy~beam$^{-1}$ in Stokes $I$.
Given the large number of resolution elements searched, the
conservative search strategies developed by \citet{wro05} were
followed.  Specifically, a 6 $\sigma$ detection threshold was adopted
and only one source emerged above this threshold.  This VLBA
detection, shown in Figure 1, is unresolved with a diameter of less
than 1.63 mas and has a flux density of $S = 0.88\pm0.12$ mJy.  The
ratio of the flux density of the VLBA detection to that measured with
the VLA \citep{wro09} is 0.75$\pm$0.11.  The position of the VLBA
detection is $\alpha(J2000) = 15^{h} 36^{m} 36\fs2232$ and
$\delta(J2000) = +04\arcdeg 41\arcmin 27\farcs069$, with a
conservative 1 $\sigma$ error of 2 mas per coordinate estimated from
analysis of the check source, J1544+0407, observed at a switching
angle of 1.5\arcdeg\, from the reference source.

We also conducted a slightly deeper (5 $\sigma$) small-area search for
an additional point-like source near the VLBA detection and within the
5400 mas$^2$ localization region described in \S~\ref{motivation} for
the broad-line emission systems.  Figure~2 shows that if an additional
source is present within that parallelogram-shaped region, it is
fainter than the VLBA detection by a factor of 2.6 or more.  This
upper limit was corrected for a 20\% coherence loss, estimated from
the ratio, about 1.2, of the VLBA detection's integrated-to-peak flux
densities.  This ratio is applicable to a switching angle of
0.7\arcdeg\, and is also consistent with the higher ratio, about 1.6,
measured for the check source which had a switching angle of
1.5\arcdeg.

\section{Implications}\label{implications}

\subsection{Size of the Radio-Sphere}\label{RS}

From the new VLA photometry, the RQQ SDSS J1536+0441A has a rising
spectrum with index $\alpha = 0.35\pm0.08$ at rest frequencies of tens
of gigahertz.  (This is consistent with our prior suggestion that the
overall spectrum of A and B was flat or rising between 1.4 GHz [White
  et al. 1997] and 8.5 GHz [Wrobel \& Laor 2009].)  SDSS J1536+0441A
thus resembles the 45\%-50\% of RQQ that show flat or rising
integrated spectra at similar frequencies \citep{bar96,ulv05}.  This
spectrum suggests that SDSS J1536+0441A is compact enough to be
synchrotron self-absorbed, as expected in the coronal framework for
RQQ \citep{lao08}.  SDSS J1536+0441A has a bolometric luminosity of
$L_{bol} = 1.5 \times 10^{46}$~ergs~s$^{-1}$ \citep{bor09} and a 22.5
GHz luminosity density of $L_R = 8.6 \times
10^{30}$~ergs~s$^{-1}$~Hz$^{-1}$.  Applying equation (22) of
\citet{lao08}, for a homogeneous synchrotron source with equipartition
between magnetic and photon energy densities, implies a radio-sphere
of radius about 0.04 pc at a rest frequency of 31.2 GHz, and thus $T_b
\sim 5 \times 10^{10}$ K.  This is just below the \citet{rea94} limit
of $T_b \sim 10^{11}$ K, expected for equipartition between the
electron energy density and the magnetic energy density within the
radio-sphere.  The 0.04 pc radius corresponds to about 800
gravitational radii for a $10^9$~${\rm M}_{\odot}$ BH, a mass thought
to be applicable to SDSS J1536+0441A \citep{lau09,tan09}.  The radius
of the radio-sphere implies that the 31.2~GHz emission arises just
within the optical broad-line emission region (BLR) discussed in
\S~\ref{BBH}.

The spectral index of this RQQ is clearly too shallow to be a
homogeneous source, and likely implies a superposition of emission
from an inhomogeneous source, as commonly adopted for flat spectrum
radio-loud systems \citep[e.g.,][]{phi85}.  Given the newly measured
flat spectral index for SDSS J1536+0441A, its ratio of radio to X-ray
\citep{arz09} luminosities drops from $5.9\times 10^{-5}$
\citep{wro09} to $1.2\times 10^{-5}$, putting it close to the
$10^{-5}$ average ratio characterizing lower-luminosity active
galactic nuclei, a ratio expected in the coronal framework for the
radio emission from RQQ \citep{lao08}.

Some RQQs are time variable \citep{bar05} so our inference of a rising
spectrum for SDSS J1536+0441A is weakened by using non simultaneous
photometry.  However, for the source to have an optically thin
spectral index steeper than $\alpha = -0.5$, the flux density at
either 8.5 or 22.5 GHz would need to vary by a factor greater than two
between the 100 days separating the measurements.  Causality arguments
would then imply a size upper limit of $100/(1+0.388) \simeq 70$ light
days, or 0.06 pc.  In this case, both the steep spectrum and the high
variability brightness temperature, $T_b > 10^{10}$ K, would exclude a
thermal free-free origin for the radio emission.

From the new VLBA imaging (Figs.~1 and 2), a single source was
detected at 8.4 GHz within the 470 pc by 310 pc localization region
for the broad-line emission (Chornock et al. 2010; T. Boroson, 2010,
private communication).  The VLBA detection has a geometric-mean
diameter of less than 8.5 pc and a rest-frame brightness temperature,
modified for an elliptical Gaussian, of $T_b > 1.2 \times 10^7$ K.
The ratio of the VLA and VLBA flux densities near 8 GHz is broadly
consistent with unity.  This suggests that the VLBA recovers all of
the VLA signal but time variability remains a concern.  For now, we
tentatively assign the VLA-derived spectral index, $\alpha =
0.35\pm0.08$, to the VLBA detection.  Then the isotropic power at a
rest frequency of 8.4 GHz is $P_\nu = 2.9 \times 10^{23}$~W~Hz$^{-1}$.
This VLBA detection has a power and $T_b$ limit at the low end of the
values reported for other RQQs detected with the VLBA
\citep{blu96,blu98,ulv05}.  The crude estimate made above for a
synchrotron-self-absorbed size is also consistent with the VLBA
detection.

The implications of the above findings are examined below, first
within the context of BBH scenarios for SDSS J1536+0441A
\citep{bor09,tan09} and then within the context of it being a lone DPE
\citep{gas10,cho10}.

\subsection{Binary Black Hole Scenarios}\label{BBH}  

Our VLBA findings are consistent with the projected separation of the
two quasars being less than 8.5 pc.  The VLBA localization area for
SDSS J1536+0441A is 82 pc$^2$, improving over the emission-line
localization area by a factor of about 1800.  The VLBA detection
appears to have a rising, synchrotron self-absorbed spectrum, which
bodes well for imaging it at higher resolutions to improve the
localization further.  Optical spectroscopic monitoring of SDSS
J1536+0441A implies an orbital period longer than about 200 years
\citep{lau09}.  Unfortunately, such a long period means that VLBA
monitoring could not usefully constrain the astrometric wobble of SDSS
J1536+0441A.

The twin DPE model of \citep{tan09} reproduces the observed H$\beta$
line profile by emission from a disk extending from 7000 down to 800
gravitational radii.  Thus, the size of the radio-sphere at 31.2 GHz
is just below the inner boundary of the BLR.  Since the size of an
optically thick radio-sphere scales as $f_\nu^{0.4}/\nu$
(e.g. equation 22 of Laor \& Behar 2008), observations at mm
wavelengths will allow us to probe the radio-sphere on smaller scales,
potentially down to the optically emitting region at a few 10s of
gravitational radii.  If the source remains optically thick, then
sub-mm observations can probe down to the X-ray emitting region at a
few gravitational radii.

The VLBA detection of the RQQ SDSS J1536+0441A represent the first
parsec-scale localization of a candidate BBH system identified through
its broad-line properties.  This VLBA detection also demonstrates that
parsec-scale localizations of candidate BBH systems need not be
restricted to radio loud objects like the radio galaxy 0402+379 with
its 7-pc separation \citep{rod06}.

Our VLBA findings cannot exclude additional sources with $T_b < 4.5
\times 10^6$ K in the same field of view.  Such a value is atypically
low compared to other RQQs detected with the VLBA
\citep{blu96,blu98,ulv05}.  But those studies targeted RQQ stronger
than several millijansky and were thus biased toward detecting higher
brightness temperatures.  Several RQQ were not detected in the survey
of \citet{blu98}, with brightness temperature limits similar to the
present study.  This suggests that the effects of source resolution
could also contribute to non detections of RQQs.

\subsection{Lone Double-Peaked Emitter Scenario}\label{DPE}

The lone DPE scenario for the RQQ SDSS J1536+0441A \citep{gas10,cho10}
requires the presence of only one quasar.  The line profiles for SDSS
J1536+0441A do make it an unusual DPE however \citep{cho10,lau09}.  As
a class, DPEs are rare, constituting only 4\% of the
spectroscopically-selected sample at $z < 0.332$ of \citet{str03}.
Yet 76\% of those DPEs are radio quiet like SDSS J1536+0441A.  Using
1.4 GHz detections from \citet{whi97}, \citet{str03} tabulate 1.4 GHz
luminosities for 18 RQ DPEs and report typical values of several times
$10^{39}$~ergs~s$^{-1}$.  SDSS J1536+0441AB is not detected by
\citet{whi97} and the 1.4 GHz luminosity is less than $7.2 \times
10^{39}$~ergs~s$^{-1}$, a limit consistent with detected RQ DPEs in
\citet{str03}.  Moreover, except for SDSS J1536+0441A, no RQQs
detected with the VLBA \citep{blu96,blu98,ulv05} are known to exhibit
DPEs, so it is impossible to say whether or not the VLBA detection of
SDSS J1536+0441A is in any way unusual.  In this regard, VLBA imaging
of the DPE sample of \citet{str03} would be a useful undertaking.  As
the VLBA detection of SDSS J1536+0441A demonstrates, such imaging is
feasible for both radio-quiet and radio-loud DPEs.

\section{Summary}\label{summary}

Our VLBA search for 8.4 GHz emission from the RQQ SDSS J1536+0441A
found only one source within the localization region, of area 0.15
kpc$^2$, for the broad-line emission.  The VLBA detection has a
diameter of less than 8.5 pc and a $T_b > 1.2 \times 10^7$ K.  This
detection of SDSS J1536+0441A represents the first parsec-scale
localization of a candidate BBH system identified through its
broad-line properties.  The VLA photometry at a rest frequency of 31.2
GHz yields a rising spectrum, consistent with synchrotron
self-absorption, which implies a radius of about 0.04 pc for the
radio-sphere, and $T_b \sim 5 \times 10^{10}$ K.  The observed compact
flat spectrum radio sphere is consistent with the trait predicted in
the coronal framework for RQQs.  The radio-sphere at 31.2 GHz happens
to be just inside the estimated inner boundary for the BLR, of 800
gravitational radii, in this object.

It would be useful to investigate the spectrum at higher frequencies,
corresponding to the mm and sub-mm range, to determine where the
spectral slope steepens as the source becomes optically thin.  This
will establish the size of the most compact synchrotron-emitting
region, and potentially allow a direct exploration of relativistic
electrons in the accretion disk corona.  We plan such investigations
using the Expanded VLA \citep{per09} and the Atacama Large
Millimeter/submillimeter Array \citep{woo09}.

Concerning the BBH scenarios for SDSS J1536+0441A, the VLBA detection
is consistent with a quasar separation of less than 8.5 pc.  No
additional sources with $T_b = 4.5 \times 10^6$ K or more are found in
within the localization region for the broad-line emission.

Concerning the lone DPE scenario for SDSS J1536+0441A, its emission
line profiles do make it an unusual DPE.  But as no other radio-quiet
DPEs have been imaged on parsec scales, it is impossible to say
whether or not the VLBA detection of SDSS J1536+0441A is in any way
unusual.

\acknowledgments We acknowledge prompt and helpful feedback from the
referee, and useful discussions with Craig Walker and Greg Taylor.
This research was supported by THE ISRAEL SCIENCE FOUNDATION (grant
\#407/08), and by a grant from the Norman and Helen Asher Space
Research Institute.

{\it Facilities:} \facility{VLA} \facility{VLBA}

\clearpage

\begin{figure}
\epsscale{1.0}
\plotone{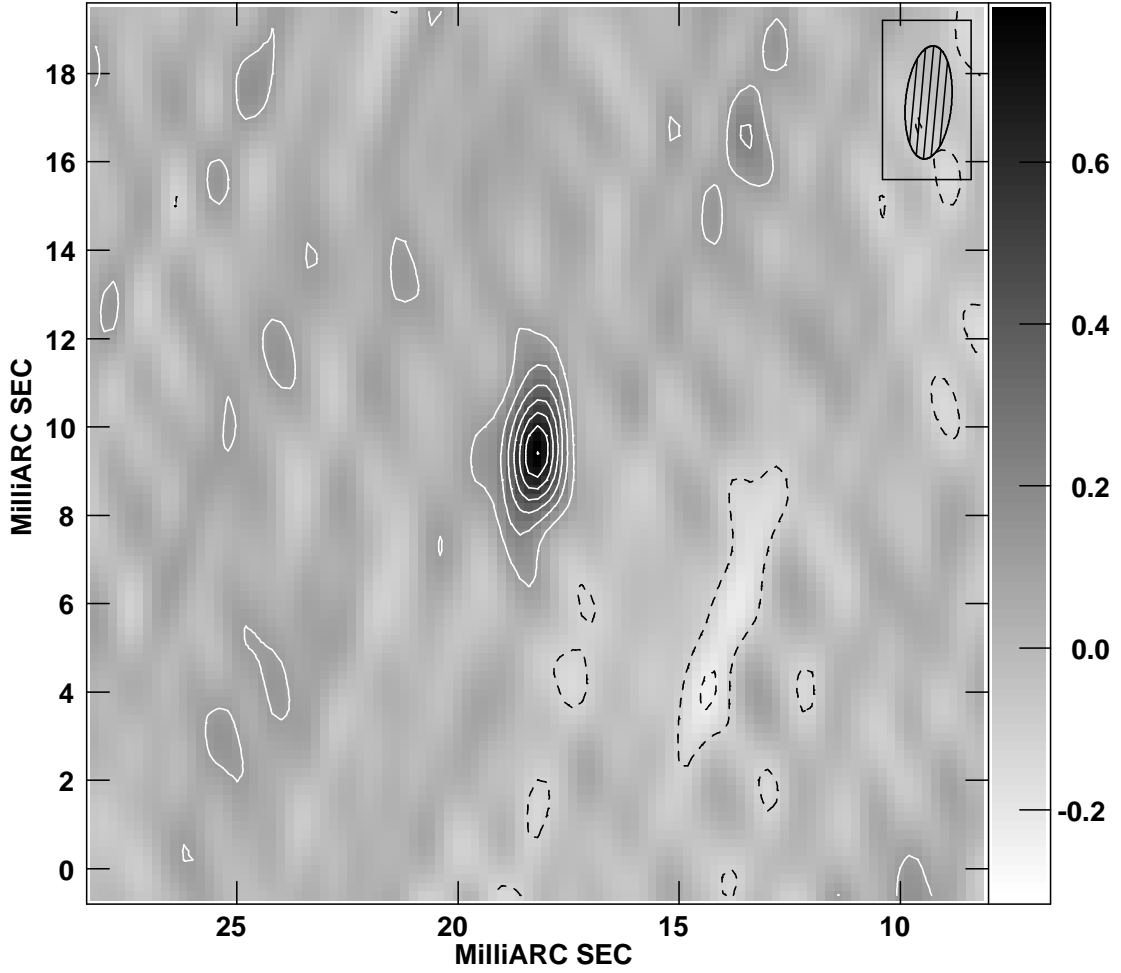}
\caption{VLBA image of Stokes $I\/$ emission from SDSS J1536+0441A at
  a frequency of 8.4~GHz and spanning 20 mas (104~pc).  The rms noise
  is 0.056~mJy~beam$^{-1}$ (1 $\sigma$) and the hatched ellipse shows
  the Gaussian beam dimensions at FWHM.  Geometric-mean beam width is
  1.63 mas (8.5 pc) at FWHM.  Contours are at $-$6, $-$4, $-$2, 2, 4,
  6, 8, 10, 12, ...  14 times 1 $\sigma$.  Negative contours are
  dashed and positive ones are solid.  Image peak is
  0.79~mJy~beam$^{-1}$.  Linear gray scale spans
  $-$0.24~mJy~beam$^{-1}$ to 0.79~mJy~beam$^{-1}$.}\label{fig1}
\end{figure}
\clearpage

\begin{figure}
\epsscale{1.0}
\plotone{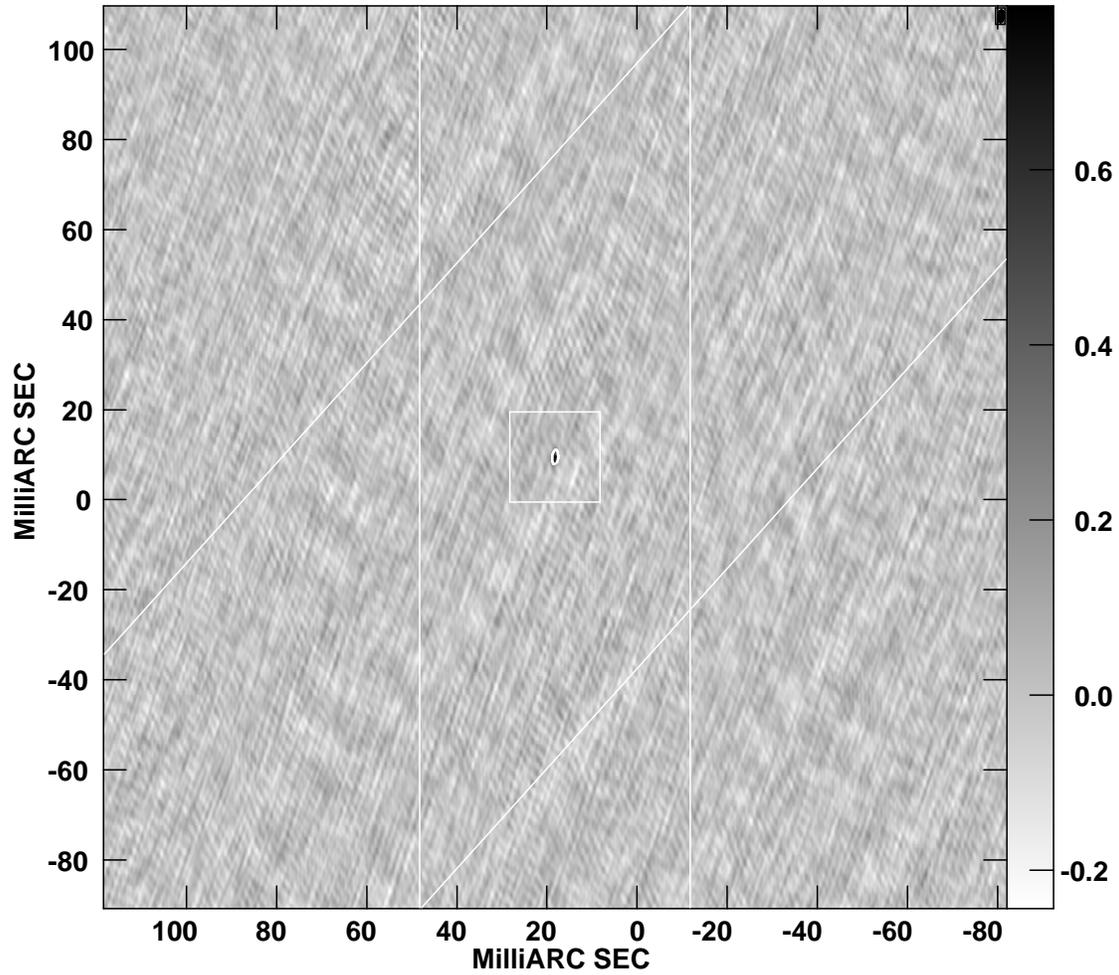}
\caption{Stokes $I\/$ emission at 8.4~GHz centered on the VLBA
  detection of SDSS J1536+0441A and spanning 200 mas (1.04~kpc). The 1
  $\sigma$ rms noise, beam dimensions, image peak and gray scale are
  as in Fig.~1.  Contours are at $-$5 and 5 times 1 $\sigma$.  The
  square shows the field of view for Fig.~1.  The parallelogram
  encloses the localization region, of area 5400 mas$^2$ (0.15
  kpc$^2$), for the broad-line emission systems.  Any additional
  point-like source within the parallelogram has an observed peak
  below 5 $\sigma$.}\label{fig2}
\end{figure}

\end{document}